\documentclass{ws-mpla}
\usepackage{slashed}
\usepackage{subfigure}

\def\EtMiss  {E_\mathrm{T}\!\!\!\!\!\!\!/ \ \; }

\begin{document}

\markboth{Wolfgang Wagner}
{Top-Quark Cross Section and Properties at the Tevatron}

\catchline{}{}{}{}{}

\title{TOP-ANTITOP-QUARK PRODUCTION AND DECAY PROPERTIES AT THE TEVATRON}

\author{\footnotesize WOLFGANG WAGNER}

\address{Fachgruppe Physik, Bergische Universit\"at Wuppertal, 
 Gau{\ss}stra{\ss}e 20, \\
42119 Wuppertal, Germany \\wagner@physik.uni-wuppertal.de}

\maketitle

\begin{center} 
CDF/PHYS/TOP/PUBLIC/10107 \\ Version 2.0 \\
FERMILAB-PUB-10-056-E \\
  \today \end{center}


\begin{abstract}
At the Tevatron, the collider experiments CDF and
  D\O \ have data sets at their disposal that comprise a few thousand
  reconstructed top-antitop-quark pairs and allow for precision 
  measurements of the cross section as well as production and decay properties.
  Besides comparing the measurements to standard model predictions,
  these data sets open a window to physics beyond the standard model.
  Dedicated analyses look for new heavy gauge bosons, fourth generation
  quarks, and flavor-changing neutral currents.
  In this mini-review the current status of these measurements is
  summarized.
\end{abstract}

\ccode{PACS Nos.: 14.65.Ha}

\section{Introduction}
The top quark is by far the heaviest elementary particle observed 
by particle physics experiments and features a mass of 
$m_t = 173.1\pm1.3\;\mathrm{GeV}/c^2$~\cite{:2009ec}.
The large mass of the top quark gives rise to large radiative 
corrections, for example to the $W$ propagator, which causes a strong 
correlation between $m_W$, $m_t$, and the Higgs boson mass $m_H$. 
To predict $m_H$ a precise measurement of $m_t$ is crucial.
The large mass leads also to a very short lifetime
of the top quark, $\tau_t \simeq 0.5\cdot 10^{-24}\,\mathrm{s}$, such that
top hadrons are not formed. The top quark thus offers the unique 
possibility to study a quasi-free quark and as a consequence 
polarization effects are accessible in the angular distributions of 
top-quark decay products. Since $m_t$ is close
to the energy scale at which the electroweak gauge symmetry breaks down 
(vacuum expectation value of the Higgs field $v=246\;\mathrm{GeV}$), 
it has been argued that the top quark may be part
of a special dynamics causing the break down of the electroweak 
gauge symmetry~\cite{peccei}.
Finally, the top quark gives access to the highest energy scales and 
offers thereby the chance to find new, unexpected physics, 
for example heavy resonances that decay into $t\bar{t}$ pairs.

In the past years the Fermilab Tevatron, a synchrotron colliding protons and 
antiprotons at a center-of-mass energy of $\sqrt{s}=1.96\;\mathrm{TeV}$, was 
the only place to produce and observe top quarks under laboratory conditions. 
Physics data taking of Tevatron Run II started in 2002 and in the meanwhile the 
accelerator has delivered collisions corresponding to an integrated luminosity 
of $8.0\;\mathrm{fb^{-1}}$. The two general-purpose detectors CDF and D\O \
have recorded collision data corresponding to $6.5\;\mathrm{fb^{-1}}$ and
$7.0\;\mathrm{fb^{-1}}$, respectively.

\section{Top-Antitop Production Cross Section}
\begin{figure}[t]
    \begin{center}
    \subfigure[]{
    \includegraphics[width=0.18\textwidth]{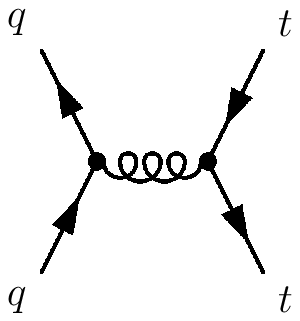}
      \label{subfig:qqtt}}  
    \hspace*{16mm}
    \subfigure[]{
    \includegraphics[width=0.18\textwidth]{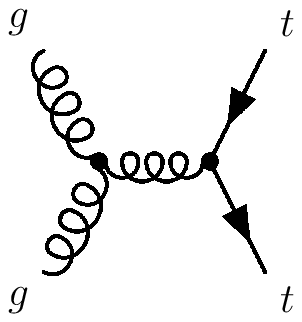}  \hspace*{4mm}
    \includegraphics[width=0.18\textwidth]{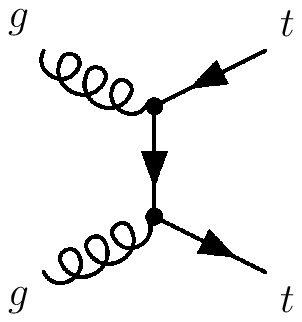}  \hspace*{4mm}
    \includegraphics[width=0.18\textwidth]{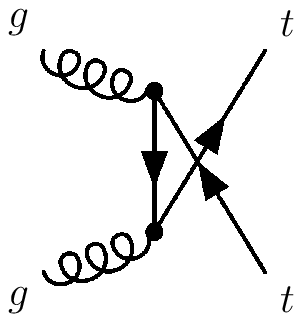}
      \label{subfig:ggtt}}
    \end{center}
    \caption{\label{fig:leadingOrderttbar} Feynman diagrams of the leading order
      processes for $t\bar{t}$ production: 
      \subref{subfig:qqtt} quark-antiquark annihilation 
      ($q\bar{q} \rightarrow t\bar{t}$) and \subref{subfig:ggtt}
      gluon-gluon fusion ($gg \rightarrow t\bar{t}$).
     }
  \end{figure}

The main source of top quarks at the Tevatron is the pair production
via the strong interaction. At leading order in perturbation theory 
($\alpha_s^2$) there are two processes that contribute to $t\bar{t}$ production,
quark-antiquark annihilation $q\bar{q} \rightarrow t\bar{t}$ and 
gluon-gluon fusion $gg \rightarrow t\bar{t}$.
The corresponding Feynman diagrams for these processes are depicted
in Fig.~\ref{fig:leadingOrderttbar}.

According to the standard model (SM) top
quarks decay with a branching ratio of nearly 100\% to a bottom quark
and a $W$ boson and the $t\bar{t}$ final states can be 
classified according to the decay modes of the $W$ bosons. The most
important (or golden) channel is the so-called {\it lepton+jets} channel
where one $W$ boson decays leptonically into a charged lepton (electron or 
muon) plus
a neutrino, while the second $W$ boson decays into jets.
The lepton+jets channel features a large branching ratio
of about 29\%, manageable backgrounds, and allows for the full 
reconstruction of the event kinematics. 
Other accessible channels are the {\it dilepton}
channel, in which both $W$ bosons decay to leptons, and the {\it all-hadronic}
channel, where both $W$ bosons decay hadronically. 
The dilepton channel has the advantage of having a low background, 
but suffers on the other hand from a lower branching fraction (5\%) 
compared to the lepton+jets channel.
The all-hadronic channel, on the contrary, has the largest branching 
ratio of all $t\bar{t}$ event categories (46\%), but 
has the drawback of a huge QCD multijet background, that has to be
controlled experimentally. The different categories of $t\bar{t}$
and their branching fractions are summarized in 
Table~\ref{tab:ttbarDecay}.
\begin{table}
  \tbl{\label{tab:ttbarDecay} Categories of $t\bar{t}$ events and their 
    branching fractions. The sum of all fractions is above 100\% because
    of rounding effects.}{
  \begin{tabular}{@{}cccc@{}}
    \toprule
    $W$ decays & $e/\mu\nu$ & $\tau\nu$ & $q\bar{q}$ \\ \hline
    $e/\mu\nu$ &   5\%     &   5\%     &    29\%    \\
    $\tau\nu$  &   --       &  1\%     &    15\%    \\
    $q\bar{q}$ &   --       &   --      &   46\%    \\ \botrule
  \end{tabular}} 
\end{table}

\subsection{Lepton+Jets Channel}
The experimental signature of lepton+jets $t\bar{t}$ events
comprises a reconstructed isolated lepton candidate, large
missing transverse energy ($\EtMiss$) and at least four jets with large
transverse energy $E_T \equiv E\cdot\sin \theta$. Two jets originate
from $b$-quarks. Typical selection cuts ask for a charged lepton with 
$p_T > 20\;\mathrm{GeV}/c$, $\EtMiss > 20\;\mathrm{GeV}$, and at least
four jets with $E_T > 20\;\mathrm{GeV}$ and $|\eta|<2.0$, one of them
identified as a $b$-quark jet. The most commonly used algorithm to 
identify $b$-quark jets is based on the reconstruction of secondary
vertices in jets, exploiting the relatively long lifetime of $b$-hadrons
and a large Lorentz boost. The typical decay length of $b$-hadrons
in high-$p_T$ $b$-quark jets is on the order of a few millimeters.
The requirement of a secondary vertex within one of the jets leads to
a large reduction of the $W$+jets background by roughly a factor
of 50, while the selection efficiency for $t\bar{t}$ events is 
about 50\% to 60\%. In Run I and Run II several analyses used 
secondary vertex $b$-jet tagging to enrich the $t\bar{t}$ signal
in the lepton+jets 
channel~\cite{Acosta:2004hw,Acosta:2004be,Abulencia:2006in,Abazov:2006ka}. 
Alternative methods identify $b$-quark jets by relying on the impact parameter 
significance of tracks associated to jets~\cite{Abazov:2005ey,Abulencia:2006kv}, 
or by reconstructing leptons originating from semileptonic decays 
of $b$ hadrons~\cite{Acosta:2005zd,:2009ax}.
Advanced $b$-jet taggers combine all available information using
neural networks~\cite{Abazov:2008gc,ttbarNNtaggerCDF}.

\begin{figure}[t]
\begin{center}
\includegraphics[width=0.7\textwidth]{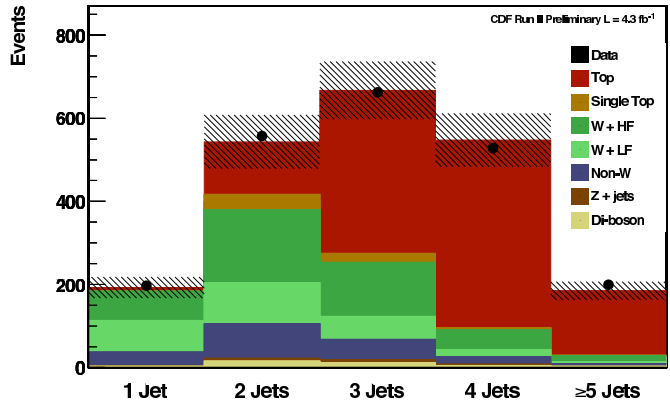}
\end{center}
\caption{\label{fig:Wjets}Jet multiplicity distribution for the 
  $W$+jets data set, where the $W$ boson is reconstructed in its leptonic 
  decay
  $W^\pm \rightarrow \ell^\pm \nu_\ell (\bar{\nu}_\ell)$. A cut on 
  $H_T>230\,\mathrm{GeV}$ was applied. The analyzed data set corresponds
  to $4.3\,\mathrm{fb^{-1}}$.} 
\end{figure}

The most recent CDF analysis based on secondary vertex 
$b$ tagging~\cite{Aaltonen:2010ic} 
is a counting experiment in which the background rate is estimated using a combination 
of simulated events and data driven methods. The signal region is
defined as the data set with a leptonic $W$ candidate plus $\geq 3$
jets. To further suppress background, a cut on the sum of all transverse
energies $H_T>230\,\mathrm{GeV}$ is applied. 
The jet multiplicity distribution of the $W$+jets data set observed by
this CDF analysis is shown in Figure~\ref{fig:Wjets}.
The uncertainty on the luminosity measurement is reduced by measuring
the ratio of $t\bar{t}$-to-$Z$-boson cross sections and the measured
cross section is found to be
$7.32\pm 0.36\,(\mathrm{stat})\pm 0.59 (\mathrm{syst})\pm 0.14\,(\mathrm{Z\;theory})\;\mathrm{pb}$, assuming $m_t = 172.5\;\mathrm{GeV}/c^2$. 
The cross section of $t\bar{t}$ production and  
$Z/\gamma^* \rightarrow \ell^+\ell-$ production are measured in 
data samples corresponding to the same integrated luminosity.
By forming the ratio of both measured cross sections and multiplying
by the well-known theoretical $Z/\gamma^* \rightarrow \ell^+\ell-$
cross section the luminosity uncertainty of 6\% is effectively removed
and replaced by the uncertainty on the  $Z/\gamma^*$ cross section
of 2\%.

The identification of $b$-quark jets is a powerful tool to remove 
$W$+jets background. However, the technique is also associated to
systematic uncertainties which become relevant when the data 
statistics increases. The uncertainties are associated to the
efficiencies to identify $b$-quark jets, but also to the efficiencies
to wrongly identify $c$-quark jets and light-quark jets as $b$ jets.
An additional uncertainty arises from the flavor composition of
the $W$+jets data set before applying the tagging algorithm.
At CDF, the fraction of heavy-flavor jets is found to be higher in 
collision data than predicted by the Monte Carlo generator 
{\sc Alpgen}~\cite{Mangano:2002ea} when studying the $W$+1 jets 
data set using a neural-network-based flavor separating 
tool~\cite{Richter:2007zzc,Renz:2008zz,Lueck:2009zz}. As a result,
the fraction of $Wb\bar{b}$ and $Wc\bar{c}$ is scaled up by a
common factor of $1.4\pm 0.4$. The large uncertainty is assigned
to cover the results of studies in higher jet-multiplicity 
samples and the results obtained by investigating other 
flavor-separating variables.

An alternative to $b$-quark jet identification is to enhance the 
$t\bar{t}$ fraction in the candidate sample by exploiting 
topological or kinematic features of $t\bar{t}$ events. These
techniques have also been used extensively at the 
Tevatron~\cite{Abazov:2005ex,Acosta:2005am,Abazov:2007kg}.

The single most precise measurement of the $t\bar{t}$ cross section
at CDF is based on a neural network technique applied to the 
$W+\geq 3\;$jets data set~\cite{ttbarNNxsCDF}. 
This method has the advantage that it does 
not use $b$-quark jet tagging and therefore avoids
the systematic uncertainties associated to the $b$-tagging efficiency
and the fraction of heavy-flavor jets in $W$+jets events. 
The neural network measurement finds a $t\bar{t}$ cross section
of $7.82\pm 0.38\,(\mathrm{stat})\pm 0.37 (\mathrm{syst})\pm 0.15\,(\mathrm{Z\;theory})\;\mathrm{pb}$ (at $m_t = 172.5\;\mathrm{GeV}/c^2$).

\subsection{Dilepton Channel}
The final state in which both $W$ bosons originating from the top quark 
decay either into $e^-\bar{\nu}_e$ or $\mu^-\bar{\nu}_\mu$ is called
{\it dilepton channel}. It features two high-$p_T$, isolated charged 
leptons and large missing transverse energy due to the undetected neutrinos.
Final states with $\tau$ leptons are generally not explicitly reconstructed
in this channel, but contribute indirectly if the $\tau$ decays 
leptonically into an electron or muon plus neutrinos. 

Measurements in the dilepton channel contributed to the discovery of the
top quark in the 1990s~\cite{Abe:1994st,Abe:1995hr,Abachi:1995iq}.
In Run II, CDF made several cross section measurements in the
dilepton channel~\cite{Acosta:2004uw,Abulencia:2006mf,Aaltonen:2009ve}.
While the standard approach requires both leptons to be well identified as 
either electrons or muons, analysts at CDF developed
a second technique that allows one of the leptons to be measured only as
a high-$p_\mathrm{T}$ isolated track, thereby significantly increasing the lepton
detection efficiency at the cost of a moderate increase in the expected 
backgrounds.

The most recent measurement by the CDF collaboration uses a data set
corresponding to $4.5\,\mathrm{fb^{-1}}$ and is based on events featuring 
a pair of oppositely charged isolated leptons with 
$p_T\geq 20\,\mathrm{GeV}/c$, $\EtMiss\geq 25\,\mathrm{GeV}$, and 
two or more jets with $E_T\geq 30\,\mathrm{GeV}$. The number of selected 
candidate events is 215 over an expected background of $66.9\pm 5.7$,
yielding a measured cross section of
$6.56\pm 0.65\,(\mathrm{stat})\pm 0.41\,(\mathrm{syst})\pm 0.38\,(\mathrm{lumi})\;\mathrm{pb}$~\cite{CDF_Note_9890} at $m_t = 175\;\mathrm{GeV}/c^2$.

\begin{figure}[t]
\begin{center}
\includegraphics[width=0.5\textwidth]{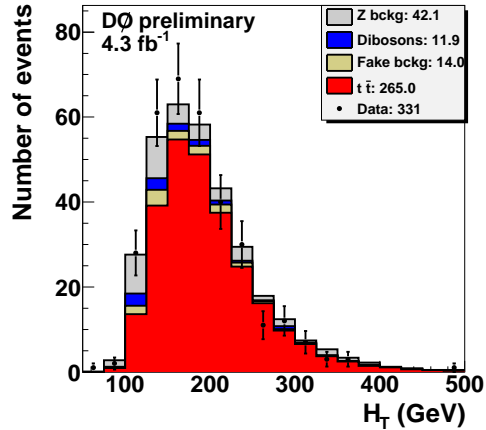}
\end{center}
\caption{\label{fig:dilepton_HT_DO}Distribution of the variable $H_T$
  for lepton+jets events of collision data corresponding to
  $4.3\,\mathrm{fb^{-1}}$ of integrated luminosity recorded
  from June 2006 to June 2009.}  
\end{figure}

At the beginning of Run II, D\O \ first used fully identified electrons and
muons for the dilepton $t\bar{t}$ cross section measurement~\cite{Abazov:2005yt},
but added later also a lepton+track event category~\cite{:2007bu}.
In a dilepton measurement with $1\,\mathrm{fb^{-1}}$ of collision data 
the D\O \ collaboration used $e\tau$ and $\mu\tau$ final states
in addition to the standard decay channels~\cite{Abazov:2009si}. The $\tau$
leptons are identified in their hadronic decay mode as a narrow jet with low
track multiplicity and the identification is optimized using neural networks.
The results of this measurement are combined with the most recent
analysis of dilepton final states using an integrated luminosity of
$4.3\,\mathrm{fb^{-1}}$.
Fig.~\ref{fig:dilepton_HT_DO} shows the distribution of the 
kinematic variable $H_T$ which is defined as the scalar sum of the
transverse momenta of the charged lepton and the jets in the event.
The resulting cross section is 
$8.4\pm 0.5\,(\mathrm{stat})^{+0.9}_{-0.8}\,(\mathrm{syst})^{+0.7}_{-0.6}\,(\mathrm{lumi})\;\mathrm{pb}$ for $m_t = 172.5\;\mathrm{GeV}/c^2$~\cite{DO_6038}.

\subsection{All-Hadronic Channel}
Despite the large branching ratio it is very challenging to isolate 
a $t\bar{t}$ signal in the all-hadronic channel because of the 
overwhelming QCD multijet background. Given the limited energy
resolution of jets the reconstruction of the $t\bar{t}$ kinematics is also
very demanding. The cross section measurements in the all-hadronic channel
therefore exploit in most cases topological or event shape information to 
discern signal from background. 
The observation of $t\bar{t}$ events in the all-hadronic channel dates back
to Run I of the Tevatron~\cite{Abe:1997rh,Abbott:1999mr}. While CDF used 
$b$-quark jet identification to suppress background, the 
D\O \ analyses employed a neural network to combine several kinematic variables
to one discriminant that separates QCD multijet background from $t\bar{t}$
events.

In Run II, CDF first published an analysis combining $b$-quark jet identification
with cuts on several topological variables~\cite{Abulencia:2006se}, but
replaced these cuts in an updated measurement by a kinematical selection
using a neural network~\cite{:2007qf}. The latest measurement in this 
series~\cite{Aaltonen:2010pe} added input variables to the neural 
network that discriminate between quark-initiated and gluon-initiated
jets. As an another novelty the analysis applied techniques used for 
the measurement of $m_t$, namely the simultaneous
determination of the top-quark mass $m_t$, the jet energy scale, and the
number of signal and background events. 
After the neural network based 
event selection the signal yields are obtained by a fit to the reconstructed
top-quark mass distribution $m_t^\mathrm{rec}$ and the mass distribution of 
the $W$ boson candidates. The resulting value for the $t\bar{t}$ cross
section is
$7.2\pm 0.5\,(\mathrm{stat})\pm 1.0\,(\mathrm{syst})\pm 0.4\,(\mathrm{lumi})\;\mathrm{pb}$ at $m_t = 172.5\;\mathrm{GeV}/c^2$.

In its first analysis of Run II the D\O \ collaboration used a combination
of secondary vertex tagging of $b$-quark jets and a neural network
to separate $t\bar{t}$ events from QCD multijet events~\cite{Abazov:2006yb}.
In an updated analysis D\O \ used an advanced $b$-tagger based on neural 
networks that features a tagging efficiency of $(57 \pm 2)\%$ for
$b$-quark jets and a misidentification rate of $(0.57 \pm 0.07)\%$.
At least two jets are required to be tagged in this way. Kinematic 
information of the events is exploited by combining several variables
to a likelihood-ratio discriminant. The measured cross section is
$6.9\pm 1.3\,(\mathrm{stat})\pm 1.4\,(\mathrm{syst})\pm 0.4\,(\mathrm{lumi})\,\mathrm{pb}$~\cite{Abazov:2009ss}, 
assuming $m_t = 175\;\mathrm{GeV}/c^2$.

\subsection{Cross Section Combination}
The measurements of the $t\bar{t}$ cross section in $p\bar{p}$ 
collisions at $\sqrt{s}=1.8\,\mathrm{TeV}$ in Run I were combined
experiment wise and gave 
$6.5^{+1.7}_{-1.4}\,\mathrm{pb}$ at CDF~\cite{Affolder:2001wd}
 (measured at $m_t = 175\;\mathrm{GeV}/c^2$) and 
$5.7\pm 1.6\,\mathrm{pb}$ at D\O~\cite{Abazov:2002gy} 
($m_t = 172.1\;\mathrm{GeV}/c^2$). The predicted  $t\bar{t}$ cross section 
at $\sqrt{s}=1.8\,\mathrm{TeV}$ is 
$5.24\pm 0.31\,\mathrm{pb}$ 
(at $m_t = 175\;\mathrm{GeV}/c^2$)~\cite{Kidonakis:2003qe}.

\begin{figure}[t]
\begin{center}
  \subfigure[]{
    \includegraphics[width=0.47\textwidth]{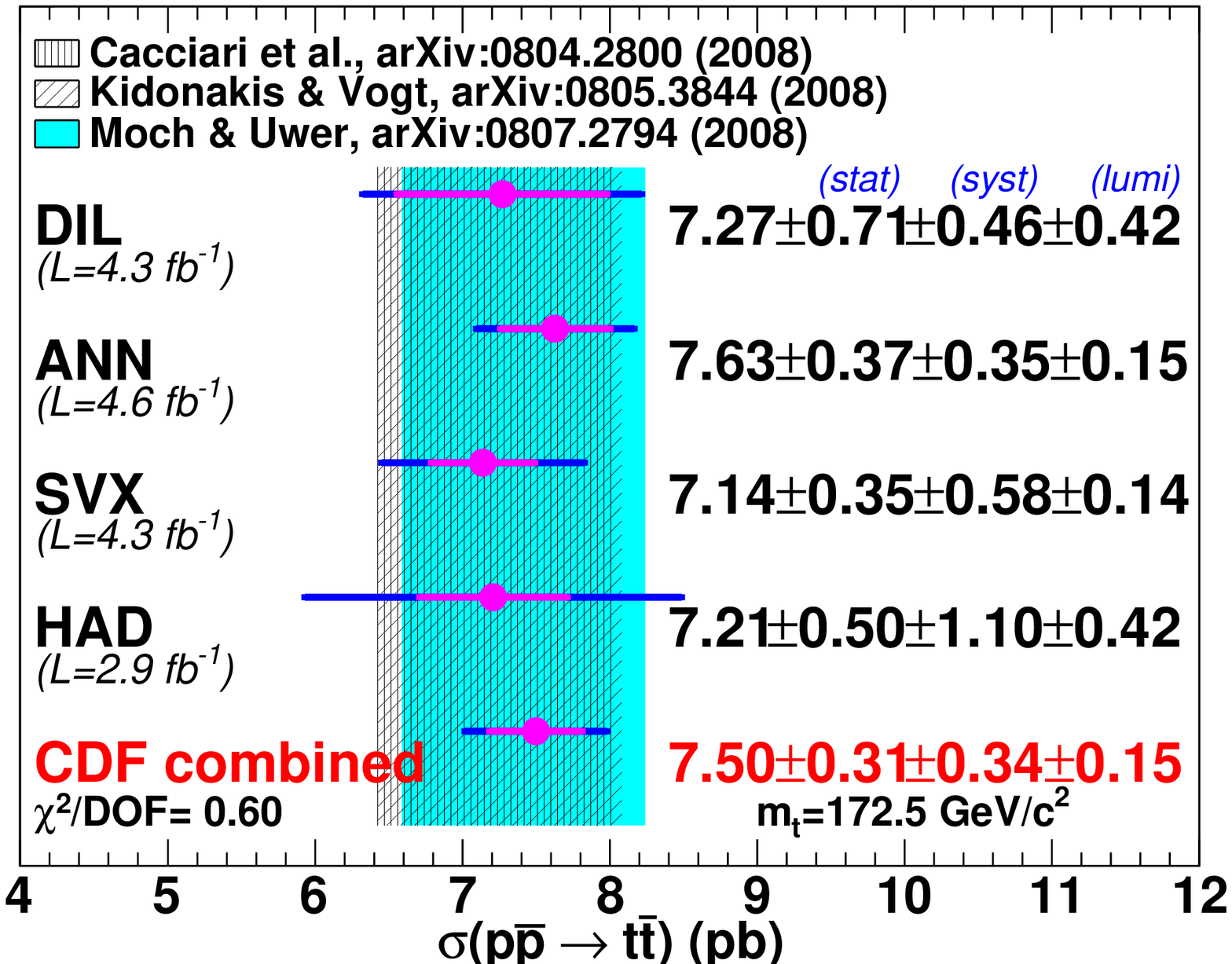}
    \label{fig:CDFcrosssections}
  }
  \subfigure[]{
    \includegraphics[width=0.48\textwidth]{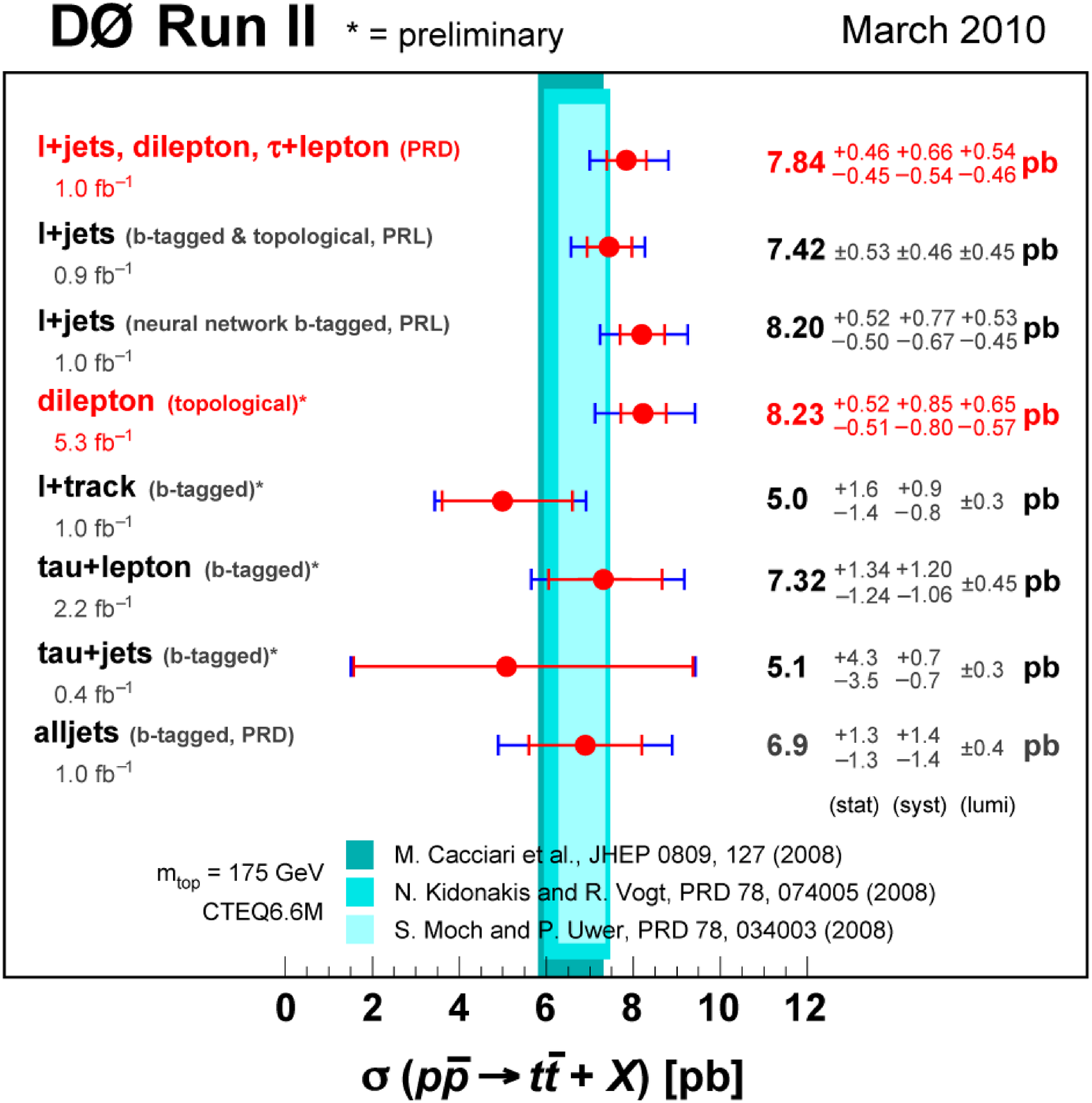}
    \label{fig:D0crosssections}
  } 
\end{center}
\caption{\label{fig:crosssections}Summary of the $t\bar{t}$ cross sections
  measured by \subref{fig:CDFcrosssections} CDF and 
  \subref{fig:D0crosssections} D\O \ compared to theoretical 
  predictions.}
\end{figure}

A summary of the $t\bar{t}$ cross section measurements at CDF in Run II 
is given in Fig.~\ref{fig:crosssections}\subref{fig:CDFcrosssections}.
The combination of these measurements yields
$7.50\pm 0.31\,(\mathrm{stat})\pm 0.34 (\mathrm{syst})\pm 0.15\,(\mathrm{Z\;theory})\;\mathrm{pb}$~\cite{CDF_Note_9913}.
The Run II cross section results of the D\O \ collaboration 
are given in 
Fig.~\ref{fig:crosssections}\subref{fig:D0crosssections}. The 
combined cross section is found to be 
$8.18^{+0.98}_{-0.87}\,\mathrm{pb}$ 
(at $m_t=170\,\mathrm{GeV}/c^2$)~\cite{Abazov:2009ae}.
All measurements are in excellent agreement with theory predictions
based on the SM~\cite{Cacciari:2008zb,Kidonakis:2008mu,Moch:2008qy} 
which give, for example, $6.90^{+0.46}_{-0.64}\,\mathrm{pb}$
at $m_t=175\,\mathrm{GeV}/c^2$~\cite{Moch:2008ai}.

\section{Production Properties}
The measured top-quark mass~\cite{:2009ec} agrees very well with the one   
predicted by the SM using electroweak precision measurements as 
input~\cite{lepewwg_2009}. Also the top-quark pair production cross 
section is in excellent agreement with the theoretical expectation.
Nevertheless, it is important to also test other predictions made by
the SM about the top quark to firmly establish its identity.

\subsection{Production Mechanism}
Calculations in perturbative QCD predict that the dominating subprocess
of $t\bar{t}$ pair production at the Tevatron is $q\bar{q}$ annihilation 
(85\%), while gluon-gluon fusion contributes 15\%. 
To measure the fraction of $t\bar{t}$ pairs originating from a $gg$
initial state, physicists at CDF exploited the fact that $gg$ initial 
states produce more initial-state radiation than $q\bar{q}$ initial states.
The analysis makes use of the proportionality of the mean number of 
low-$p_\mathrm{T}$ tracks in an event $\bar{N}_\mathrm{trk}$ and the 
gluon content. The linear relation between 
$\bar{N}_\mathrm{trk}$ and the average number of hard initial-state 
gluons is calibrated in $W$+jets and dijet data samples.
Using simulated events, templates of the $\bar{N}_\mathrm{trk}$ 
distribution are calculated for $gg\rightarrow t\bar{t}$
and $q\bar{q}\rightarrow t\bar{t}$ events. These templates are fit
to the distribution observed in collision data, resulting in 
a measurement of 
$\sigma(gg\rightarrow t\bar{t})/\sigma(q\bar{q}\rightarrow t\bar{t})=
 0.07\pm0.14\,(\mathrm{stat.})\pm 0.07\,(\mathrm{syst.})$
\cite{:2007kq}. An alternative method exploits the spin information
in the top-decay products employing neural networks and sets an upper limit
on the $gg$ initiated fraction of $t\bar{t}$ events of
0.61 at the 95\% confidence level (C.L.)~\cite{Abulencia:2008su}.

Recently, the CDF collaboration measured the $gg$-fusion fraction in 
the $t\bar{t}$ dilepton channel. The measurement exploits the fact that
$t\bar{t}$ pairs produced via $gg$ fusion are in a different spin state,
namely $J=0$, $J_z = 0$, than $t\bar{t}$ pairs produced via $q\bar{q}$
annihilation ($J=1$, $J_z = \pm 1$). The different spin state manifests
itself in different azimuthal angular correlations of the charged leptons
in $t\bar{t}$ dilepton events. Based on a data set corresponding to an
integrated luminosity of $2\,\mathrm{fb^{-1}}$ a fit to the distribution
of the difference $\Delta\phi$ of the azimuthal angles of the charged 
leptons yields a $gg$-fusion fraction of
$f_{gg}=0.53^{+0.36}_{-0.38}$~\cite{cdf9432}.
The spin correlation of $t\bar{t}$ pairs was further investigated in a
recent measurement based on an integrated luminosity of 
$4.3\,\mathrm{fb^{-1}}$~\cite{CDF10048}. 
Using lepton+jets events the fraction $f_o$ of
$t\bar{t}$ pairs in the opposite helicity state ($J=1$) was determined.
The spin correlation can be expressed by the correlation coefficient
$\kappa$ which related to $f_o$ by $f_o = \frac{1}{2}(1+\kappa)$
and found to be
$\kappa = 0.60\pm 0.50\,(\mathrm{stat})\pm 0.16\,(\mathrm{syst})$
which is in good agreement with theoretical 
calculations~\cite{Bernreuther:2001rq,Bernreuther:2004jv}.

\subsection{Forward-Backward Asymmetry}
Due to interference effects at next-to-leading order (NLO) QCD predicts 
a forward-backward asymmetry
\begin{equation}
  \label{eq:AFB_theory} 
  A_\mathrm{FB} = \frac{N_t(p) - N_{\bar{t}}(p)}{N_t(p)+ N_{\bar{t}}(p)} 
   = (5.0 \pm 1.5)\%
\end{equation}
at the Tevatron~\cite{Kuhn:1998kw,Antunano:2007da,Bernreuther:2010ny}, 
where $N_t(p)$ is the number of 
top quarks moving in proton direction and $N_{\bar{t}}(p)$ is
the number of antitop quarks moving in proton direction.
The theoretical uncertainty is driven by the size of higher order
corrections to $A_\mathrm{FB}$~\cite{Dittmaier:2007wz,Almeida:2008ug}.
The asymmetry indicates that top quarks are more likely to be produced 
in proton direction,
while antitop quarks are more likely to be produced in antiproton
direction.
The relatively small value of $A_\mathrm{FB}$ in the SM is
a net result of a positive asymmetry from the interference of the Born
amplitude with virtual box corrections ($t\bar{t}$ final state) and
a negative asymmetry from the interference of initial and final state
radiation amplitudes ($t\bar{t}g$ final state). 
While the SM value of $A_\mathrm{FB}$ is barely 
measurable at the Tevatron, the measurement is sensitive to 
non-standard model effects that can reach the size of up to $\pm 30\%$,
{\it e.g.}, in models with $Z^\prime$-like states and parity violating 
couplings~\cite{Rosner:1996eb} or theories with chiral 
gluons~\cite{Antunano:2007da}.

\begin{figure}[t]
\begin{center}
\includegraphics[width=0.7\textwidth]{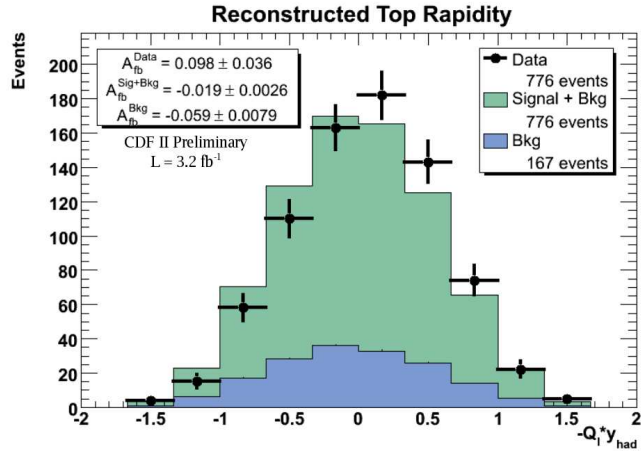}
\end{center}
\caption{\label{fig:AFB} Distribution of $Q_\ell\cdot y_\mathrm{had}$ observed
  by the CDF collaboration in collision data corresponding to an
  integrated luminosity of $3.2\,\mathrm{fb^{-1}}$. The displayed distribution is
  not corrected for acceptance and reconstruction effects, while these corrections
  are applied to obtain the measured value of $A_\mathrm{FB}$.}
\end{figure}

Using events of the lepton+jets
topology CDF and D\O \ have investigated the charge asymmetry.
In the CDF analysis the hadronic top quark is reconstructed and the
asymmetry 
\begin{equation}
  \label{eq:AFB_CDF} 
  A_\mathrm{FB}^\mathrm{lab} = \frac{N (-Q_\ell\cdot y_\mathrm{had} > 0) - 
         N (-Q_\ell\cdot y_\mathrm{had} < 0)}
        {N (-Q_\ell\cdot y_\mathrm{had} > 0) + 
         N (-Q_\ell\cdot y_\mathrm{had} < 0)}
   = 0.193 \pm 0.065 (\mathrm{stat.}) \pm 0.024 (\mathrm{sys.})
\end {equation}        
is measured~\cite{CDFafb}, where $Q_\ell$ is the charge of the lepton and
$y_\mathrm{had}$ is the rapidity of the reconstructed hadronically decaying
top quark. The CDF measurement quoted in (\ref{eq:AFB_CDF}) is corrected for 
background contributions, acceptance bias, and migration effects due 
to the reconstruction,
which has the important advantage that the measured quantity can be directly 
compared to the theoretically expected one in (\ref{eq:AFB_theory}).
The relatively large value compared to the
SM expectation confirms earlier CDF measurements~\cite{Aaltonen:2008hc}
and has a significance of about two Gaussian standard deviations.
The raw $Q_\ell\cdot y_\mathrm{had}$ distribution before the corrections
are applied is shown in Fig.~\ref{fig:AFB}.

D\O \ uses $\Delta y \equiv y_t - y_{\bar{t}}$ as an observable, applies a background 
correction and obtains $A = 0.12 \pm 0.08 \pm 0.01$~\cite{afbD0}.
To compare this value with the CDF measurements or with the theory 
prediction it has to be corrected for acceptance and migration 
effects. A prescription for this procedure is provided in ref.~\cite{afbD0}.
Based on this measurement the D\O \ collaboration derives limits on a
heavy $Z^\prime$ boson that decays to $t\bar{t}$ pairs.
  
\subsection{Top-Antitop Resonances}
The $t\bar{t}$ candidate samples offer another possibility to search for
a narrow-width resonance $X^0$ decaying into $t\bar{t}$ pairs by investigating 
the $t\bar{t}$ invariant mass~\cite{Abazov:2003aw,:2007dia}. 
In an analysis using data 
corresponding to 
$3.6\;\mathrm{fb^{-1}}$ the D\O \ collaboration found no evidence for such a 
resonance and places upper limits on 
$\sigma_X\cdot\mathrm{BR}(X^0\rightarrow t\bar{t})$ ranging from 
$1.0\,\mathrm{pb}$ at $M_X=350\,\mathrm{GeV}/c^2$ to
$0.16\,\mathrm{pb}$ at $M_X=1000\,\mathrm{GeV}/c^2$~\cite{d0Mttbar}. 
If interpreted in
the frame of a topcolor-assisted technicolor model these limits can be 
used to derive mass limits on a narrow lepto-phobic $Z^\prime$:
$M(Z^\prime) > 820\;\mathrm{GeV}/c^2$ at the 95\% C.L., assuming
$\Gamma(Z^\prime) = 0.012\,M(Z^\prime)$. A similar 
analysis in the all-hadronic channel at CDF yields slightly lower mass 
limits~\cite{CDFallhadronicTT}. 
Analysts at CDF have also searched for massive gluonic states
in the $t\bar{t}$ invariant mass spectrum, setting limits
for masses between $400$ and 
$800\,\mathrm{GeV}/c^2$~\cite{Aaltonen:2009tx}.

\subsection{Direct Searches for Fourth Generation Quarks}
A straight forward extension of the SM is the addition of a fourth 
generation of fermions. While measurements at LEP and in the 
flavor sector set stringent 
limits on the parameters of such an extension, it is not fully 
excluded 
yet~\cite{Kribs:2007nz,Hung:2007ak,Bobrowski:2009ng,Buras:2010pi}. 
The measurement of the $Z$ lineshape, for example,
implies that the fourth generation neutrino has a mass exceeding
$46.7\,\mathrm{GeV}/c^2$~\cite{Bulanov:2003ka}.
An intriguing feature of a fourth generation of quarks is the
possibility to allow for large CP violation effects that would be
big enough to explain the baryon asymmetry in the 
universe~\cite{Hou:2008xd}.

The CDF collaboration has used the $t\bar{t}$ candidate sample of 
lepton+jets events to search for 
a fourth-generation up-type quark $t^\prime$ and sets a lower limit
of $m_{t^\prime}>335\;\mathrm{GeV}/c^2$ at the 95\% 
C.L.~\cite{CDFtprime} with collision data corresponding to
$4.6\,\mathrm{pb^{-1}}$.
Another analysis searched for the pair production of heavy 
fourth-generation down-type quarks in the decay channel 
$b^\prime\bar{b}^\prime \rightarrow t W^- \bar{t} W^+$ by
looking for events with two same-charge leptons ($e$ or $\mu$),
several jets and $\EtMiss$. No significant excess of such events is
found and a lower limit of $m_{b^\prime}>338\;\mathrm{GeV}/c^2$ 
at the 95\% C.L. is set~\cite{Aaltonen:2009nr}.

\section{Top-Quark Decay Properties}
\label{sec:topdecay}
Within the SM, top quarks decay predominantly into a $b$ quark and a $W$ boson, 
while the decays $t\rightarrow d + W^+$ and $t\rightarrow s + W^+$ are strongly 
CKM suppressed and can be neglected. The top-quark decay rate, including first 
order QCD corrections, is given by
\begin{equation}
  \Gamma_{t} = \frac{G_F\;m_t^3}{8\,\pi\,\sqrt{2}}\;
  \; \left| V_{tb} \right|^2 \;
  \left( 1 - \frac{M_W^2}{m_t^2}\right)^2\;
  \left( 1 + 2\; \frac{M_W^2}{m_t^2}\right)\;
  \left[ 1 - \frac{2\,\alpha_s}{3\,\pi} \cdot 
  f\left(y\right)
  \right]
  \label{eq:topdecay}
\end{equation}
with $y=(M_W/m_t)^2$ and 
$f(y) = 2\,\pi^2/3-2.5-3y+4.5y^2-3y^2\ln{y}$~\cite{Jezabek:1987nf,Jezabek:1988iv,Kuhn:1996ug},
yielding $\Gamma_t =1.34\;\mathrm{GeV}$ at $m_t = 172.6\;\mathrm{GeV}/c^2$. 
The large $\Gamma_t$ implies a very short lifetime of 
$\tau_t = 1/\Gamma_t \approx 0.5\cdot10^{-24}\,\mathrm{s}$ 
which is smaller than the characteristic formation time of hadrons 
$\tau_\mathrm{form} \approx 1\,\mathrm{fm}/c\approx 3\cdot 10^{-24}\;\mathrm{s}$.
In other words, top quarks decay before they can couple to light quarks and 
form hadrons. 

\subsection{W-Boson Helicity in Top-Quark Decays}
The amplitude of the decay $t\rightarrow b + W^+$ is dominated by the 
contribution from longitudinal $W$ bosons because the decay rate of the 
longitudinal component scales with $m_t^3$, while the decay rate into 
transverse $W$ bosons increases only linearly with $m_t$. In both cases the 
$W^+$ couples solely to $b$ quarks of left-handed chirality, which translates 
into left-handed helicity, since the $b$ quark is effectively massless compared 
to the energy scale set by $m_t$. If the $b$ quark is emitted antiparallel to 
the top-quark spin axis, the $W^+$ must be longitudinally polarized, 
$h^W$ = 0, to conserve angular momentum.
If the $b$ quark is emitted parallel to the top-quark spin axis, the $W^+$ boson
has helicity $h^W = -1$ and is transversely polarized. 
$W$ bosons with positive helicity are thus forbidden in top-quark decays due to 
angular momentum conservation, assuming $m_b=0$.
The fraction of longitudinally polarized $W$ bosons is predicted to be 
$f_0=m_t^2/(2 m_W^2+m_t^2) \simeq 0.70$ and the left-handed fraction
$f_{-}=2 m_W^2/(2 m_W^2+m_t^2)\simeq 0.30$.
Taking the non-zero $b$-quark mass into account yields very small 
corrections, leading to a non-zero fraction of $W$ bosons in top-quark decay 
with positive helicity 
$f_+ = 3.6\cdot 10^{-4}$ at Born level~\cite{Fischer:2000kx}. 

\begin{figure}[t]
\begin{center}
\includegraphics[width=0.6\textwidth]{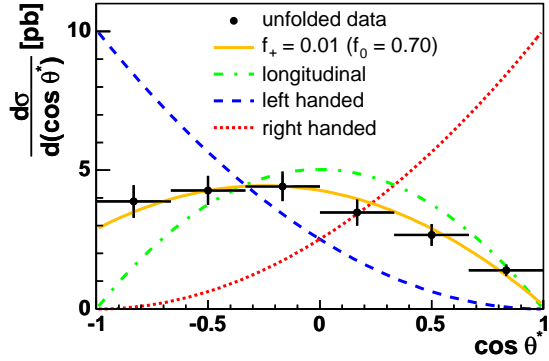}
\end{center}
\caption{\label{fig:Whel} The deconvoluted distribution of 
  $\cos\theta^*$ normalized to the inclusive
  $t\bar{t}$ cross section is compared to theoretically predicted curves
  of purely left handed, right handed, and longitudinally polarized
  $W$ bosons from top-quark decay. The measurement by the CDF collaboration 
  uses data corresponding to $1.9\,\mathrm{fb^{-1}}$.}
\end{figure}

The spin orientation (helicity) of $W$ bosons from top-quark decays is 
propagated to its decay products. In case of leptonic $W$ decays the polarization 
is preserved and can be measured. A useful observable for experiments is therefore
the cosine of the polarization angle $\theta^*$ which is defined as the  
angle of the charged lepton in the $W$-boson decay frame measured with respect to 
the top-quark direction. The probability density $\omega$ has the following 
distribution:
\begin{eqnarray}
\label{eq:omega}
\omega (\theta^{*}) = { f_{0} \cdot \omega_0 (\theta^{*}) +
                   { f_{+} \cdot \omega_+ (\theta^{*}) +
                   { (1-f_{0} -f_{+}) \cdot \omega_- (\theta^{*}) }}} 
  \ \ \mathrm{with} \\ 
\omega_0 (\theta^{*}) = \frac{3}{4}(1-\cos^{2}\theta^{*}), \ 
\omega_+ (\theta^{*}) = \frac{3}{8}(1+\cos\theta^{*})^{2}, \
\omega_- (\theta^{*}) = \frac{3}{8}(1-\cos\theta^{*})^{2}. 
\end{eqnarray}
\noindent
Several measurements of the helicity fractions have been performed by the 
CDF and D\O \  collaborations. One class of analyses reconstruct the $t\bar{t}$
event kinematics in lepton$+$jets events and measure $\cos\theta^*$ 
directly~\cite{Abazov:2005fk,Abazov:2006hb,Abulencia:2006ei,Abazov:2007ve,Aaltonen:2008ei},
while a second class uses the square of the invariant mass of the lepton and the 
$b$-quark jet $M_{\ell b}^2$ in the laboratory frame as 
observable~\cite{Acosta:2004mb,Abulencia:2005xf,Abulencia:2006iy}. 
The variable $M_{\ell b}^2$ is closely 
related to $\cos\theta^*$ and has the advantage that it can also be used
in $t\bar{t}$ dilepton events.
In Fig.~\ref{fig:Whel} the observed distribution of 
$\cos\theta^*$~\cite{Aaltonen:2008ei} after deconvoluting acceptance and 
reconstruction effects is compared to the probability densities $\omega$
of the three $W$ boson helicities scaled to the $t\bar{t}$ cross section.

A third set of analyses~\cite{Abulencia:2005xf,Affolder:1999mp} uses the $p_T$ 
spectrum of the charged lepton which is also sensitive to the helicity 
fractions because the $V$-$A$ structure of the $W$-boson decay causes 
a strong correlation between the helicity of the $W$ boson and the 
lepton momentum. Qualitatively, this can be understood as follows:
The $\nu_\ell$ from the $W^+$ decay is always left-handed, the 
$\ell^+$ is right-handed. In the case of a left-handed $W^+$ boson angular 
momentum conservation demands therefore that the $\ell^+$ is emitted
in the direction of the $W^+$ spin, that means antiparallel to the 
$W^+$ momentum. That is why charged leptons from the decay of 
left-handed $W$ bosons are softer than charged leptons from longitudinal
$W$ bosons, which are mainly emitted in the direction transverse to the
$W$ boson momentum. 
The spectrum of leptons from right-handed $W$ bosons would be even 
more harder than the one from longitudinal ones, 
since they would be emitted preferentially in the direction
of the $W$ momentum.

A fourth technique to measure the helicity fractions uses the LO matrix
elements of $t\bar{t}$ 
production~\cite{Abazov:2004ym,CDF_CONF_9144,Aaltonen:2010ha}.

More recent analyses measure the fractions $f_0$ and $f_+$ simultaneously
in a two-dimensional fit, while in previous analyses two fits were 
performed keeping one of the fractions at its SM 
value~\cite{Abazov:2007ve,Aaltonen:2008ei,Aaltonen:2010ha}. 
Using the matrix element technique CDF obtains
$f_0 = 0.88\pm 0.11\,\mathrm{(stat)}\pm 0.06\,\mathrm{(syst)}$
and
$f_+ = -0.15\pm 0.07\,\mathrm{(stat)}\pm 0.06\,\mathrm{(syst)}$
with a correlation coefficient of $-59\%$~\cite{Aaltonen:2010ha}. 
The result is in good agreement with the SM prediction.

\subsection{Form Factors of the Wtb Vertex}
The D\O \ collaboration has performed a very general analysis of the
$Wtb$ vertex~\cite{Abazov:2009ky}. When including operators up to
dimension five, the Dirac structure of the $Wtb$ vertex can be 
generalized by the interaction Lagrangian
\begin{equation}
  \label{eq:Wtb}
  \mathcal{L} = \frac{g_w}{\sqrt{2}} \left[ W_\mu^- \bar{b}\gamma^\mu
  (f_1^L P_- + f_1^R P_+ ) t - \frac{1}{m_W} \partial_\nu W_\mu^- \bar{b}
  \sigma^{\mu\nu} (f_2^L P_- + f_2^R P_+) t \right] + h.c.\;,
\end{equation}
where $P_\pm = \frac{1}{2}(1\pm\gamma^5)$ and 
$i\sigma^{\mu\nu}=-\frac{1}{2}[\gamma^\mu, \gamma^\nu]$~\cite{Kane:1991bg}.
The four form factors $f_{1,2}^{L,R}$ are assumed to be real numbers.
In standard electroweak theory they take the values $f_1^L=1$ and 
$f_1^R=f_2^L=f_2^R=0$, such that the production of right-handed $W$ bosons 
from top-quark decay is suppressed. 
D\O \ pursues a general strategy to experimentally determine allowed regions
of parameter space for all four form factors in Eq.~\ref{eq:Wtb}. 
The experimental input is the distribution of $\cos\theta^*$ as observed
in $t\bar{t}$ candidate events and the rate of single top-quark 
events~\cite{Abazov:2009ky}.

\subsection{The measurement of $\mathcal{R}_b$}
\label{subsec:top_Rb}
In the SM the top quark is predicted to decay to a $W$ boson and a $b$ quark
with a branching fraction 
$\mathcal{R}_b \equiv \mathrm{BR}(t\rightarrow Wb)$ close to 100\%.
This prediction is obtained in the following way.
In general, the top quark can decay in three channels 
$t\rightarrow d/s/b +W^+$ and $\mathcal{R}_b$ is given by the ratio of 
the squares of the relevant CKM matrix elements: 
$\mathcal{R}_b = |V_{tb}|^2/(|V_{td}|^2+|V_{ts}|^2+|V_{tb}|^2$.
In the SM the CKM matrix has to be unitary 
($\mathbf{V V^\dag} = \mathbf{V^\dag V} = \mathbf{1}$), which 
leads to $|V_{td}|^2+|V_{ts}|^2+|V_{tb}|^2 = 1$ and thereby to
$\mathcal{R}_b = |V_{tb}|^2$.
Our present knowledge on $|V_{tb}|$ stems primarily from measurements 
of $b$-meson and $c$-meson decays which determine the values of the 
other CKM matrix elements. Using the unitarity condition of the CKM matrix 
one can obtain $|V_{tb}|$ in an indirect way. 
This method yields
$|V_{tb}| = 0.999133 \pm 0.000044$ with very high precision~\cite{Amsler:2008zzb}. 

Only recently a determination of $|V_{tb}|$ without unitarity assumption
was obtained from the measurement of the single top-quark cross section,
yielding $|V_{tb}|=0.88\pm0.07$~\cite{Group:2009qk}.
However, if a fourth generation of quarks was present, the unitarity of 
the $3\times 3$ CKM matrix could be violated. Therefore, it is desirable to 
make a direct measurement of $\mathcal{R}_b$ using $t\bar{t}$ candidate events.

In most $t\bar{t}$ cross section analyses the assumption $\mathcal{R}_b=1$ is 
made, but CDF and D\O \ have also made two measurements without this 
constraint~\cite{Affolder:2000xb,Acosta:2005hr,Abazov:2008yn}. 
In the latest analysis from D\O~\cite{Abazov:2008yn} the $W$+jets data set is split 
in various disjoint 
subsets according to the number
of jets (0, 1, or $\geq 2$), the charged lepton type (electron or muon),
and most importantly the number of $b$-tagged jets. The fit results are:
$\mathcal{R}_b = 0.97^{+0.09}_{-0.08}$ and 
$\sigma(t\bar{t}) = 8.18^{+0.90}_{-0.84} \pm 0.50\;(\mathrm{lumi})\,\mathrm{pb}$,
where the statistical and systematic uncertainties have been combined.
The lower limit on $\mathcal{R}_b$ is determined to be 
$\mathcal{R}_b > 0.79$ at the 95\% C.L.

\subsection{Search for Non-SM Top-Quark Decays}
Two different classes of non-SM top-quark decays have been searched for at the
Tevatron, decays via flavor-changing neutral currents (FCNC) and into a charged
Higgs boson.
\subsubsection{FCNC-induced Top-Quark Decays}
In the SM FCNC are not present at tree level, but rather occur only through 
loop processes in higher orders of perturbation theory. In the top sector in 
particular FCNC are strongly suppressed with branching ratios of 
$\mathcal{O} \approx 10^{-14}$. The CDF collaboration has searched for
non-SM top quark decays of the type
$t\bar{t}\rightarrow ZqW^-\bar{b}\rightarrow(\ell^+\ell^-q)(q\bar{q}^\prime\bar{b})$. 
Top-antitop candidates are reconstructed in events with four
high-$p_\mathrm{T}$ jets and two isolated leptons using kinematic
constraints. Within their experimental resolutions determined from
simulated events the reconstructed mass of the hadronically decaying 
$W$-boson $M_{qq}$ has to be equal to $m_W$, 
the mass of the reconstructed hadronic top quark $M_{bqq}$ and 
the mass of semileptonically decaying top quark $M_{Zq}$ has to resemble
$m_t$. A $\chi^2$ formed from these conditions is used to determine
the most likely combination of physics objects and measure its
$t\bar{t}$-likeness under the anomalous-decay hypothesis.
A fit to the $\sqrt{\chi^2}$ distribution in combination with a 
Feldman-Cousins technique is used to derive an upper limit on the
branching ratio: $\mathrm{BR}(t\rightarrow Zq)< 3.7\%$ at the 95\% 
C.L.~\cite{:2008aaa}. A previous Run I result by CDF also set a limit
on FCNC branching ratios in the photon-plus-jet mode:
$\mathrm{BR}(t\rightarrow \gamma q) < 3.2\%$~\cite{Abe:1997fz}.

\subsubsection{Top-Quark Decays to a Charged Higgs Boson}
In the SM a single complex Higgs doublet scalar field is responsible for
breaking the electroweak symmetry and generating the masses of gauge
bosons and fermions. Many extensions of the SM include a Higgs sector
with two Higgs doublets and are therefore called Two Higgs Doublet 
Models (THDM). In a THDM electroweak symmetry breaking leads to five physical
Higgs bosons: two neutral $CP$-even scalars $h^0$ and $H^0$, one neutral $CP$-odd
pseudoscalar $A^0$, and a pair of charged scalars $H^\pm$.
The parameters of the extended Higgs sector include 
the mass of the charged Higgs $M_{H^+}$ and $\tan\beta=v_1/v_2$,
the ratio of the vacuum expectation values $v_1$ and $v_2$ of the 
two Higgs doublets. The minimal supersymmetric model (MSSM) is an
example of a THDM.

If the charged Higgs boson is lighter than the difference between top-quark 
mass and $b$-quark mass, $m_{H^\pm}<m_{t}-m_b$, the decay mode 
$t \rightarrow H^+ b$ is possible and competes with the SM decay
$t\rightarrow W^+b$. The branching fraction depends on $\tan\beta$ and
$m_{H^+}$. The MSSM predicts that the channel $t \rightarrow H^+ b$
dominates the top-quark decay for $\tan\beta \gtrsim 70$.
In most analyses it is assumed that 
$\mathrm{BR}(t\rightarrow W^+b)+\mathrm{BR}(t\rightarrow H^+b) = 1$.
In the parameter region $\tan\beta < 1$ the dominant decay mode is
$H^+\rightarrow c\bar{s}$, while for $\tan\beta > 1$ the decay channel
$H^+\rightarrow \tau^+\nu_\tau$ is the most important one.
For $\tan\beta > 5$ the branching fraction to $\tau^+\nu_\tau$ is nearly
100\%. Thus, in this region of parameter space THDM models predict 
an excess of $t\bar{t}$ events with tau leptons over the SM expectation.

First searches for the $H^\pm$ in top-quark events were already performed 
well before the top quark was discovered in 1994/95. The UA1 and UA2 
experiments at the CERN $Sp\bar{p}S$ excluded certain regions of the
$m_{t}$ versus $m_{H^\pm}$ 
plane~\cite{Albajar:1990zs,Alitti:1992hv}. 
In Run I of the Tevatron, CDF and D\O \  improved these limits using 
events with a dilepton signature~\cite{Abe:1994ng} or reconstructing 
tau leptons in their hadronic decay 
mode~\cite{Abe:1993mr,Abe:1995pj,Abe:1997rk,Affolder:1999au,Abazov:2001md}, which 
is experimentally a very challenging task at a hadron collider.
In Tevatron Run II, CDF has added a direct search for a charged Higgs boson
in the decay channel $H^+\rightarrow c\bar{s}$ using the dijet invariant
mass as a discriminant~\cite{Aaltonen:2009ke}. In a reanalysis of the 
$t\bar{t}$ candidate samples used for cross section measurements CDF also
considered the decay modes $H^+\rightarrow t^*\bar{b}$ and 
$H^+\rightarrow W^+h^0$ with $h^0 \rightarrow b\bar{b}$~\cite{Abulencia:2005jd}.
The D\O \ collaboration has recently performed direct searches in the 
lepton$+$jets~\cite{Abazov:2009wy} and dilepton channels~\cite{:2009zh}, 
but has also done a simultaneous measurement of the
$t\bar{t}$ cross section in the lepton$+$jets, the dilepton and the
$\tau$$+$lepton channel, deriving in parallel limits on top-quark decays 
to a charged Higgs boson~\cite{Abazov:2009ae}. 
So far, there is no evidence for $t \rightarrow H^+ b$ decays.
The resulting limits on the branching ratio $\mathrm{BR}(t\rightarrow H^+b)$ depend on 
$m_{H^\pm}$ and 
assumptions made on the Higgs decays modes, but are typically in the range 
of 15\% to 25\%.

\section{Conclusions}
Large data sets of $t\bar{t}$ candidate events are now available to the 
Tevatron collaborations CDF and D\O \ and facilitate precise investigations 
of top-quark properties.
The $t\bar{t}$ production cross section has been measured with a relative 
precision of 6.5\% and is in excellent agreement with QCD 
predictions~\cite{Moch:2008ai}. 
Many interesting analyses have searched for physics beyond the SM, for 
example for resonances decaying into $t\bar{t}$ pairs. 
The measured forward-backward asymmetry in $t\bar{t}$ production
shows an intriguing excess which will come under closer scrutiny once more
data will be analysed.
The measurements of top-quark decay show impressive 
progress, for example the measurement of the $W$-helicity fractions in
top decay.
More detailed reviews on top-quark physics are available on
phenomenology~\cite{Bernreuther:2008ju},
Tevatron Run I 
results~\cite{Campagnari:1996ai,Bhat:1998cd,Tollefson:1999wt,Chakraborty:2003iw},
and Tevatron Run 
II 
results~\cite{Wagner:2005jh,Quadt:2007jk,Kehoe:2007px,Pleier:2008ig,Incandela:2009pf,Heinson:2010xh}. 

\section*{Acknowledgments}
The author would like to thank his colleagues Frederic Deliot, 
Fabrizio Margaroli, and Tom Schwarz as well as Werner Bernreuther 
for useful discussions on the
manuscript and acknowledges the financial support of the 
Helmholtz-Alliance {\it Physics at the Terascale}.


\begin{thebibliography}{0}
\bibitem{:2009ec}
    [Tevatron Electroweak Working Group and CDF Collaboration and D0 Collab],
  arXiv:0903.2503 [hep-ex].

\bibitem{peccei} R.D. Peccei, S. Peris, and X. Zhang, Nucl. Phys. B 349
  (1991) 305--322.

\bibitem{Acosta:2004hw}
  D.~E.~Acosta {\it et al.}  [CDF Collaboration],
  Phys.\ Rev.\  D {\bf 71} (2005) 052003
  [arXiv:hep-ex/0410041].

\bibitem{Acosta:2004be}
  D.~E.~Acosta {\it et al.}  [CDF-II Collaboration],
  Phys.\ Rev.\  D {\bf 71} (2005) 072005
  [arXiv:hep-ex/0409029].

\bibitem{Abulencia:2006in}
  A.~Abulencia {\it et al.}  [CDF Collaboration],
  Phys.\ Rev.\ Lett.\  {\bf 97} (2006) 082004
  [arXiv:hep-ex/0606017].

\bibitem{Abazov:2006ka}
  V.~M.~Abazov {\it et al.}  [D0 Collaboration],
  Phys.\ Rev.\  D {\bf 74} (2006) 112004
  [arXiv:hep-ex/0611002].

\bibitem{Abazov:2005ey}
  V.~M.~Abazov {\it et al.}  [D0 Collaboration],
  Phys.\ Lett.\  B {\bf 626} (2005) 35
  [arXiv:hep-ex/0504058].

\bibitem{Abulencia:2006kv}
  A.~Abulencia {\it et al.}  [CDF Collaboration and CDF - Run II
                  Collaboration],
  Phys.\ Rev.\  D {\bf 74} (2006) 072006
  [arXiv:hep-ex/0607035].

\bibitem{Acosta:2005zd}
  D.~E.~Acosta {\it et al.}  [CDF Collaboration],
  Phys.\ Rev.\  D {\bf 72} (2005) 032002
  [arXiv:hep-ex/0506001].

\bibitem{:2009ax}
  T.~Aaltonen {\it et al.}  [CDF Collaboration],
  Phys.\ Rev.\  D {\bf 79} (2009) 052007
  [arXiv:0901.4142 [hep-ex]].

\bibitem{Abazov:2008gc}
  V.~M.~Abazov {\it et al.}  [D0 Collaboration],
  Phys.\ Rev.\ Lett.\  {\bf 100} (2008) 192004
  [arXiv:0803.2779 [hep-ex]].

\bibitem{ttbarNNtaggerCDF} T. Aaltonen {\it et al.} [CDF Collaboration],
  public conf. note no. 10049, January 2010.

\bibitem{Aaltonen:2010ic}
  T.~Aaltonen {\it et al.}  [CDF Collaboration],
  arXiv:1004.3224 [hep-ex].


\bibitem{Mangano:2002ea}
  M.~L.~Mangano, M.~Moretti, F.~Piccinini, R.~Pittau and A.~D.~Polosa,
  JHEP {\bf 0307} (2003) 001
  [arXiv:hep-ph/0206293].

\bibitem{Richter:2007zzc}
  S.~Richter, Ph.D. thesis (University of Karlsruhe),
  FERMILAB-THESIS-2007-35 (2007). 

\bibitem{Renz:2008zz}
  M.~Renz, Master thesis (University of Karlsruhe),
  FERMILAB-MASTERS-2008-06 (2008).

\bibitem{Lueck:2009zz}
  J.~Lueck, Ph.D. thesis (University of Karlsruhe),
  FERMILAB-THESIS-2009-33 (2009).

\bibitem{Abazov:2005ex}
  V.~M.~Abazov {\it et al.}  [D0 Collaboration],
  Phys.\ Lett.\  B {\bf 626} (2005) 45
  [arXiv:hep-ex/0504043].

\bibitem{Acosta:2005am}
  D.~E.~Acosta {\it et al.}  [CDF Collaboration],
  Phys.\ Rev.\  D {\bf 72} (2005) 052003
  [arXiv:hep-ex/0504053].

\bibitem{Abazov:2007kg}
  V.~M.~Abazov {\it et al.}  [D0 Collaboration],
  Phys.\ Rev.\  D {\bf 76} (2007) 092007
  [arXiv:0705.2788 [hep-ex]].

\bibitem{ttbarNNxsCDF} T. Aaltonen {\it et al.} (CDF Collaboration),
  public conf. note no. 9474, August 2009.



\bibitem{Abe:1994st}
  F.~Abe {\it et al.}  [CDF Collaboration],
  Phys.\ Rev.\  D {\bf 50} (1994) 2966.

\bibitem{Abe:1995hr}
  F.~Abe {\it et al.}  [CDF Collaboration],
  Phys.\ Rev.\ Lett.\  {\bf 74} (1995) 2626
  [arXiv:hep-ex/9503002].

\bibitem{Abachi:1995iq}
  S.~Abachi {\it et al.}  [D0 Collaboration],
  Phys.\ Rev.\ Lett.\  {\bf 74} (1995) 2632
  [arXiv:hep-ex/9503003].

\bibitem{Acosta:2004uw}
  D.~E.~Acosta {\it et al.}  [CDF Collaboration],
  Phys.\ Rev.\ Lett.\  {\bf 93} (2004) 142001
  [arXiv:hep-ex/0404036].

\bibitem{Abulencia:2006mf}
  A.~Abulencia {\it et al.}  [CDF Collaboration],
  Phys.\ Rev.\  D {\bf 78} (2008) 012003
  [arXiv:hep-ex/0612058].

\bibitem{Aaltonen:2009ve}
  T.~Aaltonen {\it et al.}  [CDF Collaboration],
  Phys.\ Rev.\  D {\bf 79} (2009) 112007
  [arXiv:0903.5263 [hep-ex]].

\bibitem{CDF_Note_9890}
  T.~Aaltonen {\it et al.}  [CDF Collaboration],
  public conf. note no. 9890, August 2009. 

\bibitem{Abazov:2005yt}
  V.~M.~Abazov {\it et al.}  [D0 Collaboration],
  Phys.\ Lett.\  B {\bf 626} (2005) 55
  [arXiv:hep-ex/0505082].

\bibitem{:2007bu}
  V.~M.~Abazov {\it et al.}  [D0 Collaboration],
  Phys.\ Rev.\  D {\bf 76} (2007) 052006
  [arXiv:0706.0458 [hep-ex]].

\bibitem{Abazov:2009si}
  V.~M.~Abazov {\it et al.}  [D0 Collaboration],
  Phys.\ Lett.\  B {\bf 679} (2009) 177
  [arXiv:0901.2137 [hep-ex]].

\bibitem{DO_6038}
  V.~M.~Abazov {\it et al.}  [D0 Collaboration],
   public conf. note no. 6038, March 2010.

\bibitem{Abe:1997rh}
  F.~Abe {\it et al.}  [CDF Collaboration],
  Phys.\ Rev.\ Lett.\  {\bf 79} (1997) 1992.

\bibitem{Abbott:1999mr}
  B.~Abbott {\it et al.}  [D0 Collaboration],
  Phys.\ Rev.\ Lett.\  {\bf 83} (1999) 1908
  [arXiv:hep-ex/9901023].

\bibitem{Abulencia:2006se}
  A.~Abulencia {\it et al.}  [CDF - Run II Collaboration],
  Phys.\ Rev.\  D {\bf 74} (2006) 072005
  [arXiv:hep-ex/0607095].

\bibitem{:2007qf}
  T.~Aaltonen {\it et al.}  [CDF Collaboration],
  Phys.\ Rev.\  D {\bf 76} (2007) 072009
  [arXiv:0706.3790 [hep-ex]].

\bibitem{Aaltonen:2010pe}
  T.~Aaltonen {\it et al.}  [The CDF Collaboration],
  arXiv:1002.0365 [hep-ex]. Accepted by Phys. Rev. D.

\bibitem{Abazov:2006yb}
  V.~M.~Abazov {\it et al.}  [D0 Collaboration],
  Phys.\ Rev.\  D {\bf 76} (2007) 072007
  [arXiv:hep-ex/0612040].

\bibitem{Abazov:2009ss}
  V.~M.~Abazov {\it et al.}  [D0 Collaboration],
  arXiv:0911.4286 [hep-ex].

\bibitem{Affolder:2001wd}
  A.~A.~Affolder {\it et al.}  [CDF Collaboration],
  Phys.\ Rev.\  D {\bf 64} (2001) 032002
  [Erratum-ibid.\  D {\bf 67} (2003) 119901]
  [arXiv:hep-ex/0101036].

\bibitem{Abazov:2002gy}
  V.~M.~Abazov {\it et al.}  [D0 Collaboration],
  Phys.\ Rev.\  D {\bf 67} (2003) 012004
  [arXiv:hep-ex/0205019].

\bibitem{Kidonakis:2003qe}
  N.~Kidonakis and R.~Vogt,
  Phys.\ Rev.\  D {\bf 68} (2003) 114014
  [arXiv:hep-ph/0308222].

\bibitem{CDF_Note_9913}
  T.~Aaltonen {\it et al.}  [CDF Collaboration],
  public conf. note no. 9913, October 2009. 

\bibitem{Abazov:2009ae}
  V.~M.~Abazov {\it et al.}  [D0 Collaboration],
  Phys.\ Rev.\  D {\bf 80} (2009) 071102
  [arXiv:0903.5525 [hep-ex]].

\bibitem{Cacciari:2008zb}
  M.~Cacciari, S.~Frixione, M.~L.~Mangano, P.~Nason and G.~Ridolfi,
  JHEP {\bf 0809} (2008) 127
  [arXiv:0804.2800 [hep-ph]].

\bibitem{Kidonakis:2008mu}
  N.~Kidonakis and R.~Vogt,
  Phys.\ Rev.\  D {\bf 78} (2008) 074005
  [arXiv:0805.3844 [hep-ph]].

\bibitem{Moch:2008qy}
  S.~Moch and P.~Uwer,
  Phys.\ Rev.\  D {\bf 78} (2008) 034003
  [arXiv:0804.1476 [hep-ph]].

\bibitem{Moch:2008ai}
  S.~Moch and P.~Uwer,
  Nucl.\ Phys.\ Proc.\ Suppl.\  {\bf 183} (2008) 75
  [arXiv:0807.2794 [hep-ph]].

\bibitem{lepewwg_2009}LEP and Tevatron Electroweak Working Groups,
  CERN-PH-EP/2009-023, FERMILAB-TM-2446-E, November 2009. 

\bibitem{:2007kq}
  T.~Aaltonen {\it et al.}  [CDF Collaboration],
  Phys.\ Rev.\  D {\bf 78}, 111101 (2008)
  [arXiv:0712.3273 [hep-ex]].

\bibitem{Abulencia:2008su}
  T.~Aaltonen {\it et al.}  [CDF Collaboration],
  Phys.\ Rev.\  D {\bf 79}, 031101 (2009)
  [arXiv:0807.4262 [hep-ex]].

\bibitem{cdf9432}
   T.~Aaltonen {\it et al.}  [CDF Collaboration],
   CDF public note no. 9432, July 2008. 

\bibitem{CDF10048}
   T.~Aaltonen {\it et al.}  [CDF Collaboration],
   CDF public note no. 10048, January 2010. 

\bibitem{Bernreuther:2001rq}
  W.~Bernreuther, A.~Brandenburg, Z.~G.~Si and P.~Uwer,
  Phys.\ Rev.\ Lett.\  {\bf 87} (2001) 242002
  [arXiv:hep-ph/0107086].

\bibitem{Bernreuther:2004jv}
  W.~Bernreuther, A.~Brandenburg, Z.~G.~Si and P.~Uwer,
  Nucl.\ Phys.\  B {\bf 690} (2004) 81
  [arXiv:hep-ph/0403035].

\bibitem{Kuhn:1998kw}
  J.~H.~Kuhn and G.~Rodrigo,
  Phys.\ Rev.\  D {\bf 59} (1999) 054017
  [arXiv:hep-ph/9807420].

\bibitem{Antunano:2007da}
  O.~Antunano, J.~H.~Kuhn and G.~Rodrigo,
  Phys.\ Rev.\  D {\bf 77} (2008) 014003
  [arXiv:0709.1652 [hep-ph]].

\bibitem{Bernreuther:2010ny}
  W.~Bernreuther and Z.~G.~Si,
  arXiv:1003.3926 [hep-ph].

\bibitem{Dittmaier:2007wz}
  S.~Dittmaier, P.~Uwer and S.~Weinzierl,
  Phys.\ Rev.\ Lett.\  {\bf 98} (2007) 262002
  [arXiv:hep-ph/0703120].

\bibitem{Almeida:2008ug}
  L.~G.~Almeida, G.~Sterman and W.~Vogelsang,
  Phys.\ Rev.\  D {\bf 78} (2008) 014008
  [arXiv:0805.1885 [hep-ph]].

\bibitem{Rosner:1996eb}
  J.~L.~Rosner,
  Phys.\ Lett.\  B {\bf 387} (1996) 113
  [arXiv:hep-ph/9607207].

\bibitem{CDFafb} T.~Aaltonen {\it et al.}  [CDF Collaboration],
  public conf. note no. 9724, March 2009.

\bibitem{afbD0}V.M. Abazov {\it et al.} [D\O \ Collaboration],
  Phys. Rev. Lett. 100, 142002 (2008).

\bibitem{Aaltonen:2008hc}
  T.~Aaltonen {\it et al.}  [CDF Collaboration],
  Phys.\ Rev.\ Lett.\  101, 202001 (2008)
  [arXiv:0806.2472 [hep-ex]].

\bibitem{Abazov:2003aw}
  V.~M.~Abazov {\it et al.}  [D0 Collaboration],
  Phys.\ Rev.\ Lett.\  {\bf 92} (2004) 221801
  [arXiv:hep-ex/0307079].

\bibitem{:2007dia}
  T.~Aaltonen {\it et al.}  [CDF Collaboration],
  Phys.\ Rev.\  D {\bf 77} (2008) 051102
  [arXiv:0710.5335 [hep-ex]].

\bibitem{d0Mttbar}V.M. Abazov {\it et al.} [D\O \ Collaboration],
  public conf. note no. 5882 (2009).

\bibitem{CDFallhadronicTT} T.~Aaltonen {\it et al.}  [CDF Collaboration],
  public conf. note no. 9844, July 2009.

\bibitem{Aaltonen:2009tx}
  T.~Aaltonen {\it et al.}  [CDF Collaboration],
  arXiv:0911.3112 [hep-ex].

\bibitem{Kribs:2007nz}
  G.~D.~Kribs, T.~Plehn, M.~Spannowsky and T.~M.~P.~Tait,
  Phys.\ Rev.\  D {\bf 76} (2007) 075016
  [arXiv:0706.3718 [hep-ph]].

\bibitem{Hung:2007ak}
  P.~Q.~Hung and M.~Sher,
  Phys.\ Rev.\  D {\bf 77} (2008) 037302
  [arXiv:0711.4353 [hep-ph]].

\bibitem{Bobrowski:2009ng}
  M.~Bobrowski, A.~Lenz, J.~Riedl and J.~Rohrwild,
  Phys.\ Rev.\  D {\bf 79} (2009) 113006
  [arXiv:0902.4883 [hep-ph]].

\bibitem{Buras:2010pi}
  A.~J.~Buras, B.~Duling, T.~Feldmann, T.~Heidsieck, C.~Promberger and S.~Recksiegel,
  arXiv:1002.2126 [hep-ph].

\bibitem{Bulanov:2003ka}
  S.~S.~Bulanov, V.~A.~Novikov, L.~B.~Okun, A.~N.~Rozanov and M.~I.~Vysotsky,
  Phys.\ Atom.\ Nucl.\  {\bf 66} (2003) 2169
  [Yad.\ Fiz.\  {\bf 66} (2003) 2219]
  [arXiv:hep-ph/0301268].

\bibitem{Hou:2008xd}
  W.~S.~Hou,
  Chin.\ J.\ Phys.\  {\bf 47} (2009) 134
  [arXiv:0803.1234 [hep-ph]].

\bibitem{CDFtprime} T.~Aaltonen {\it et al.}  [CDF Collaboration],
  public conf. note no. 10110, March 2010. 

\bibitem{Aaltonen:2009nr}
  T.~Aaltonen {\it et al.}  [The CDF Collaboration],
  arXiv:0912.1057 [hep-ex].

\bibitem{Jezabek:1987nf}
  M.~Jezabek and J.~H.~Kuhn,
  Phys.\ Lett.\  B {\bf 207} (1988) 91.

\bibitem{Jezabek:1988iv}
  M.~Jezabek and J.~H.~Kuhn,
  Nucl.\ Phys.\  B {\bf 314} (1989) 1.

\bibitem{Kuhn:1996ug}
  J.~H.~Kuhn,
  arXiv:hep-ph/9707321.

\bibitem{Fischer:2000kx}
  M.~Fischer, S.~Groote, J.~G.~Korner and M.~C.~Mauser,
  Phys.\ Rev.\  D {\bf 63} (2001) 031501
  [arXiv:hep-ph/0011075].

\bibitem{Abazov:2005fk}
  V.~M.~Abazov {\it et al.}  [D0 Collaboration],
  Phys.\ Rev.\  D {\bf 72} (2005) 011104
  [arXiv:hep-ex/0505031].

\bibitem{Abazov:2006hb}
  V.~M.~Abazov {\it et al.}  [D0 Collaboration],
  Phys.\ Rev.\  D {\bf 75} (2007) 031102
  [arXiv:hep-ex/0609045].

\bibitem{Abulencia:2006ei}
  A.~Abulencia {\it et al.}  [CDF II Collaboration],
  Phys.\ Rev.\  D {\bf 75} (2007) 052001
  [arXiv:hep-ex/0612011].

\bibitem{Abazov:2007ve}
  V.~M.~Abazov {\it et al.}  [D0 Collaboration],
  Phys.\ Rev.\ Lett.\  {\bf 100} (2008) 062004
  [arXiv:0711.0032 [hep-ex]].

\bibitem{Aaltonen:2008ei}
  T.~Aaltonen {\it et al.}  [CDF Collaboration],
  Phys.\ Lett.\  B 674, 160 (2009)
  [arXiv:0811.0344 [hep-ex]].

\bibitem{Acosta:2004mb}
  D.~E.~Acosta {\it et al.}  [CDF Collaboration],
  Phys.\ Rev.\  D {\bf 71} (2005) 031101
  [Erratum-ibid.\  D {\bf 71} (2005) 059901]
  [arXiv:hep-ex/0411070].

\bibitem{Abulencia:2005xf}
  A.~Abulencia {\it et al.}  [CDF Collaboration],
  Phys.\ Rev.\  D {\bf 73} (2006) 111103
  [arXiv:hep-ex/0511023].

\bibitem{Abulencia:2006iy}
  A.~Abulencia {\it et al.}  [CDF Collaboration],
  Phys.\ Rev.\ Lett.\  {\bf 98} (2007) 072001
  [arXiv:hep-ex/0608062].

\bibitem{Affolder:1999mp}
  A.~A.~Affolder {\it et al.}  [CDF Collaboration],
  Phys.\ Rev.\ Lett.\  {\bf 84} (2000) 216
  [arXiv:hep-ex/9909042].

\bibitem{Abazov:2004ym}
  V.~M.~Abazov {\it et al.}  [D0 Collaboration],
  Phys.\ Lett.\  B {\bf 617} (2005) 1
  [arXiv:hep-ex/0404040].

\bibitem{CDF_CONF_9144} T. Aaltonen {\it et al.} [CDF Collaboration],
  public conf. note no. 9144, November 2007.

\bibitem{Aaltonen:2010ha}
  T.~Aaltonen {\it et al.}  [The CDF Collaboration],
  arXiv:1003.0224 [hep-ex].

\bibitem{Abazov:2009ky}
  V.~M.~Abazov {\it et al.}  [D0 Collaboration],
  Phys.\ Rev.\ Lett.\  {\bf 102} (2009) 092002
  [arXiv:0901.0151 [hep-ex]].

\bibitem{Kane:1991bg}
  G.~L.~Kane, G.~A.~Ladinsky and C.~P.~Yuan,
  Phys.\ Rev.\  D {\bf 45} (1992) 124.

\bibitem{Amsler:2008zzb}
  C.~Amsler {\it et al.}  [Particle Data Group],
  Phys.\ Lett.\  B {\bf 667} (2008) 1.

\bibitem{Group:2009qk}
  T.~E.~W.~Group  [CDF Collaboration and D0 Collaboration],
  arXiv:0908.2171 [hep-ex].

\bibitem{Affolder:2000xb}
  A.~A.~Affolder {\it et al.}  [CDF Collaboration],
  Phys.\ Rev.\ Lett.\  {\bf 86} (2001) 3233
  [arXiv:hep-ex/0012029].

\bibitem{Acosta:2005hr}
  D.~E.~Acosta {\it et al.}  [CDF Collaboration],
  Phys.\ Rev.\ Lett.\  {\bf 95} (2005) 102002
  [arXiv:hep-ex/0505091].

\bibitem{Abazov:2008yn}
  V.~M.~Abazov {\it et al.}  [D0 Collaboration],
  Phys.\ Rev.\ Lett.\  {\bf 100} (2008) 192003
  [arXiv:0801.1326 [hep-ex]].

\bibitem{:2008aaa}
  T.~Aaltonen {\it et al.}  [CDF Collaboration],
  Phys.\ Rev.\ Lett.\  {\bf 101} (2008) 192002
  [arXiv:0805.2109 [hep-ex]].

\bibitem{Abe:1997fz}
  F.~Abe {\it et al.}  [CDF Collaboration],
  Phys.\ Rev.\ Lett.\  {\bf 80} (1998) 2525.

\bibitem{Albajar:1990zs}
  C.~Albajar {\it et al.}  [UA1 Collaboration],
  Phys.\ Lett.\  B {\bf 257} (1991) 459.

\bibitem{Alitti:1992hv}
  J.~Alitti {\it et al.}  [UA2 Collaboration],
  Phys.\ Lett.\  B {\bf 280} (1992) 137.

\bibitem{Abe:1994ng}
  F.~Abe {\it et al.}  [CDF Collaboration],
  Phys.\ Rev.\ Lett.\  {\bf 73} (1994) 2667.

\bibitem{Abe:1993mr}
  F.~Abe {\it et al.}  [CDF Collaboration],
  Phys.\ Rev.\ Lett.\  {\bf 72} (1994) 1977.

\bibitem{Abe:1995pj}
  F.~Abe {\it et al.}  [CDF Collaboration],
  Phys.\ Rev.\  D {\bf 54} (1996) 735
  [arXiv:hep-ex/9601003].

\bibitem{Abe:1997rk}
  F.~Abe {\it et al.}  [CDF Collaboration],
  Phys.\ Rev.\ Lett.\  {\bf 79} (1997) 357
  [arXiv:hep-ex/9704003].

\bibitem{Affolder:1999au}
  A.~A.~Affolder {\it et al.}  [CDF Collaboration],
  Phys.\ Rev.\  D {\bf 62} (2000) 012004
  [arXiv:hep-ex/9912013].

\bibitem{Abazov:2001md}
  V.~M.~Abazov {\it et al.}  [D0 Collaboration],
  Phys.\ Rev.\ Lett.\  {\bf 88} (2002) 151803
  [arXiv:hep-ex/0102039].

\bibitem{Aaltonen:2009ke}
  T.~Aaltonen {\it et al.}  [CDF Collaboration],
  Phys.\ Rev.\ Lett.\  {\bf 103} (2009) 101803
  [arXiv:0907.1269 [hep-ex]].

\bibitem{Abulencia:2005jd}
  A.~Abulencia {\it et al.}  [CDF Collaboration],
  Phys.\ Rev.\ Lett.\  {\bf 96} (2006) 042003
  [arXiv:hep-ex/0510065].

\bibitem{Abazov:2009wy}
  V.~M.~Abazov {\it et al.}  [D0 Collaboration],
  Phys.\ Rev.\  D {\bf 80} (2009) 051107
  [arXiv:0906.5326 [hep-ex]].

\bibitem{:2009zh}
  V.~M.~Abazov {\it et al.}  [D0 Collaboration],
  Phys.\ Lett.\  B {\bf 682} (2009) 278
  [arXiv:0908.1811 [hep-ex]].

\bibitem{Bernreuther:2008ju}
  W.~Bernreuther,
  J.\ Phys.\ G {\bf 35} (2008) 083001
  [arXiv:0805.1333 [hep-ph]].

\bibitem{Campagnari:1996ai}
  C.~Campagnari and M.~Franklin,
  Rev.\ Mod.\ Phys.\  {\bf 69} (1997) 137
  [arXiv:hep-ex/9608003].

\bibitem{Bhat:1998cd}
  P.~C.~Bhat, H.~Prosper and S.~S.~Snyder,
  Int.\ J.\ Mod.\ Phys.\  A {\bf 13} (1998) 5113
  [arXiv:hep-ex/9809011].

\bibitem{Tollefson:1999wt}
  K.~Tollefson and E.~W.~Varnes,
  Ann.\ Rev.\ Nucl.\ Part.\ Sci.\  {\bf 49} (1999) 435.

\bibitem{Chakraborty:2003iw}
  D.~Chakraborty, J.~Konigsberg and D.~L.~Rainwater,
  Ann.\ Rev.\ Nucl.\ Part.\ Sci.\  {\bf 53} (2003) 301
  [arXiv:hep-ph/0303092].


\bibitem{Wagner:2005jh}
  W.~Wagner,
  Rept.\ Prog.\ Phys.\  {\bf 68} (2005) 2409
  [arXiv:hep-ph/0507207].

\bibitem{Quadt:2007jk}
  A.~Quadt,
  Eur.\ Phys.\ J.\  C {\bf 48} (2006) 835.

\bibitem{Kehoe:2007px}
  R.~Kehoe, M.~Narain and A.~Kumar,
  Int.\ J.\ Mod.\ Phys.\  A {\bf 23} (2008) 353
  [arXiv:0712.2733 [hep-ex]].

\bibitem{Pleier:2008ig}
  M.~A.~Pleier,
  Int.\ J.\ Mod.\ Phys.\  A {\bf 24} (2009) 2899
  [arXiv:0810.5226 [hep-ex]].

\bibitem{Incandela:2009pf}
  J.~R.~Incandela, A.~Quadt, W.~Wagner and D.~Wicke,
  Prog.\ Part.\ Nucl.\ Phys.\  {\bf 63} (2009) 239
  [arXiv:0904.2499 [hep-ex]].

\bibitem{Heinson:2010xh}
  A.~P.~Heinson  [CDF and D0 Collaboration],
  Mod.\ Phys.\ Lett.\  A {\bf 25} (2010) 309
  [arXiv:1002.4167 [hep-ex]].

\end{thebibliography}
\end{document}